\documentclass[aps,pre,twocolumn,superscriptaddress,showpacs]{revtex4-1}
\usepackage{graphicx}
\usepackage{CJK}
\usepackage[utf8]{inputenc}
\newcommand{\bea}{\begin{eqnarray}}
\newcommand{\eea}{\end{eqnarray}}
\newcommand{\rvec}{\vec{r}}
\newcommand{\vvec}{\vec{v}}

\newcommand{\Fvec}{\vec{F}}
\newcommand{\xivec}{\vec{\xi}}
\newcommand{\f}{\frac}
\newcommand{\sgn}{\text{sgn}}

\begin{document}
\begin{CJK}{UTF8}{gbsn}

\title{Keeping speed and distance for aligned motion}

\author{Ill\'{e}s J. Farkas}
\email{fij@elte.hu}
\affiliation{
\leftline{MTA-ELTE Statistical and Biological Physics Research Group (Hungarian Academy of Sciences), P\'azm\'any P\'eter s\'et\'any 1A,}
\leftline{\,\,\,Budapest, Hungary, 1117}
}
\affiliation{\leftline{Regional Knowledge Center, ELTE Faculty of Sciences, Ir\'anyi D\'aniel u. 4., Sz\'ekesfeh\'erv\'ar, Hungary, 8000}}
\author{Jeromos Kun}
\affiliation{\leftline{Department of Biological Physics, E\"otv\"os University, P\'azm\'any P\'eter s\'et\'any 1A, Budapest, Hungary, 1117}}
\author{Yi Jin (金\hspace{1pt}\raisebox{-0.15ex}{\includegraphics[scale=0.08]{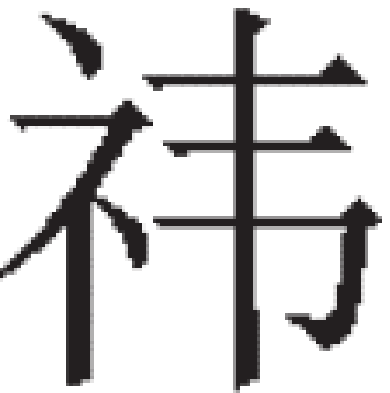}}\hspace{0.5pt})} 
%
\author{Gaoqi He (何高奇)}
\affiliation{
\leftline{Department of Computer Science and Engineering, East China University of Science and Technology, 130 Meilong Rd,}
\leftline{\,\,\,Shanghai, China, 200237}
}
\author{Mingliang Xu (徐明亮)}
\affiliation{\leftline{School of Information Engineering, Zhengzhou University, 100 Science Avenue, Zhengzhou, China, 450001}}

%
\begin{abstract}
The cohesive collective motion (flocking, swarming) of autonomous agents
is ubiquitously observed and exploited in both natural and man-made settings,
thus, minimal models for its description are essential.
In a model with continuous space and time we find that if two particles arrive symmetrically in a plane at a large angle, then
(i) radial repulsion and
(ii) linear self-propelling toward a fixed preferred speed
are sufficient for them to depart at a smaller angle.
For this local gain of momentum explicit velocity alignment is not necessary,
nor are adhesion/attraction, inelasticity or anisotropy of the particles, or nonlinear drag.
With many particles obeying these microscopic rules of motion 
we find that their spatial confinement to a square with periodic boundaries 
(which is an indirect form of attraction)
leads to stable macroscopic ordering.
%
%
As a function of the strength of added noise we see 
-- at finite system sizes --
a critical slowing down close to the order-disorder boundary and a discontinuous transition.
After varying the density of particles at constant system size and 
varying the size of the system with constant particle density
we predict that in the infinite system size (or density)
limit the hysteresis loop disappears and the transition becomes continuous.
%
%
We note that animals, humans, drones, etc. tend to move asynchronously and are often more responsive to motion than positions.
Thus, for them velocity-based continuous models can provide higher precision than coordinate-based models.
An additional characteristic and realistic
feature of the model is that convergence to the ordered state is fastest at a finite density,
which is in contrast to models applying 
(discontinuous) explicit velocity alignments 
and discretized time.
In summary, we find that the investigated model can provide a minimal description of flocking.
%
\end{abstract}

\pacs{02.70.Ns, 05.65.+b, 64.60.Cn, 87.10.Ed}
\maketitle
\end{CJK}


\section{Introduction:\\
Self-propelled motion of\\
interacting particles}
\label{s_intro}

Over the past decades {\it hypothesis-driven modeling}
has become a major method of discovery well beyond
the traditional boundaries of quantitative science.
Modeling tools and concepts from the natural sciences (including physics) and engineering
are now routinely applied to a broad range of biological, social, economic and even communication
systems \cite{alon,marchetti,castellano,mantegna,reynolds,olfati,rps}.
Clearly, this ongoing shift is also largely due to improved experimental and computational technologies.
With these tools and approaches both experimentalists and theoreticians found
that interacting cells, animals or humans often benefit from aligning their actions \cite{rainey,parrish,axelrod}.
The sum of such aligned (correlated) actions in a group is often called a group behavior.
One of the most frequently observed and best known group behaviors is {\it collective motion in biology}
when, for example, cells, insects, birds or fish move in stable spatiotemporal patterns \cite{cisneros,couzin,bonabeau,nagy10,portugal,hemelrijk}.
In addition to biological cases at the cellular or animal level,
collective motion phenomena are also common in physics, information technology, robotics and the
social sciences \cite{kudrolli,buerkle,brambilla,tarcai,helbing}.

The {\it model components} most frequently applied when describing collective motion
are (i) nonzero particle speed, 
(ii) instantaneous explicit velocity alignment (also called neighbor following)
and (iii) the discretization of simulation time.
Nonzero particle speed is necessary for modeling,
because, for example, birds/fish have to move in the air/water and humans (pedestrians) do not like to wait indefinitely.
Explicit velocity alignment is a simple, yet highly fruitful, model component.
From the physicist's point of view
its groundbreaking novelty stems from merging a self-propelled particle's ``spin'' (a pseudovector)
with its velocity (a real vector) \cite{vicsek95}.
As for the discretization of simulation time, its main purpose is to speed up computation,
which is achieved mainly through the synchronous update of all particles.
However, the discretization of time makes low particle speeds necessary to avoid artefacts
caused by the periodic boundaries \cite{nagy07,aldana}.
Moreover, discretized time combined with explicit velocity alignment
requires large enough time steps to avoid the non-physically high accelerations
that are caused by full velocity alignments within single simulation updates.

Recently, high-resolution lightweight sensors
have largely improved observations of the collective motion patterns
of autonomous moving agents \cite{nagy10,portugal,gonzalez}.
With the growth of the amount and quality of experimental data on
the many forms of collective motion it is becoming clear that there
is no single combination of model components to uniformly describe 
the large variety of flocking (swarming) phenomena
\cite{desai,jadbabaie,buhl,schweitzer,ishikawa,cavagna,ginelli,sumino,vicsek-zafeiris,swarmanoid}.
Nonetheless, the efficient prediction, de novo construction 
and control of collective motion
requires the identification of {\it minimal models},
which may also provide a classification of observed behaviors.
%
%
The need for minimal models is underlined by
the remarkable fact that despite the diversity of disciplines where collective motion has been reported
{\it several results} have been found to be {\it broadly valid}.
One such result is that with explicit velocity alignment,
high particle density and low noise the system has one stable state, and in this state rotational symmetry is broken:
all particles move in the same direction (with small random deviations) \cite{vicsek95,toner95and98,jadbabaie,gregoire,cucker-smale}.

At high densities collective motion has to slow down,
therefore, models continuous in time become more relevant \cite{couzin,helbing}.
Here we investigate stable directional alignment (ordering) during the motion of interacting particles
in a simple particle-based minimal model that is continuous in both space and time.
The model contains only the three essential features of collective motion:
particles (i) move and (ii) avoid collisions with (iii) some noise.
First, 
we find 
that in a symmetric planar collision of two identical particles the total momentum usually grows.
%
Next, we investigate whether symmetry breaking on the microscopic level
can lead to a global symmetry breaking: ordering in a 2d system of many interacting particles.
We find indeed stable ordering, measure the stability of the ordered state as a function of the noise amplitude,
and investigate the type of the transition.

\section{Model: Radial repulsion and\\
Self-propelling parallel to velocity}
\label{s_model}

We work with a model discussed in Vicsek and Zafeiris \cite{vicsek-zafeiris}
based on previous unpublished research by Derzsi, Sz\"{o}ll\H osi and Vicsek.
%
%
The model is continuous in both time and space,
and contains $N$ interacting self-propelled point-like particles.
The $i$. particle's position is $\rvec_i(t)$,
its velocity is $\vvec_i(t)$
and its self-propelling force adjusts $|\vvec_i|$ to 
a constant $v_0$ with a characteristic time of $\tau$.
The $i$. and $j$. particles interact through the radial repulsion term $\Fvec(|\rvec_i-\rvec_j|)$.
We set the mass of each particle to $m=1kg$ and apply the uncorrelated noise term,
$\xivec_i(t)$, that has a random direction and constant $\xi$ magnitude.
(The numerical implementation of the noise term is described in the caption of Fig.\,\ref{figPow}.)
In summary, the $i$. particle's equation of motion is

\bea
m\f{d \vvec_i}{dt} = m\f{\vvec_i}{|\vvec_i|} \, \f{v_0-|\vvec_i|}{\tau} - \sum_{j\not= i}\Fvec(|\rvec_i-\rvec_j|) + \xivec_i(t) \, . \,\,
\label{eq1}
\eea

In the classical ($\tau\to\infty$) limit there is no self-propelling,
thus, the system's total momentum, total angular momentum
and total (kinetic plus potential) energy are all conserved.
In the overdamped ($\tau\to 0$) limit all speeds are constant ($v_0$), thus,
the kinetic energy of each particle is conserved, but
the system's total energy, momentum and angular momentum are not conserved.
For finite $\tau$ these quantities are not conserved.

\section{Results: Ordering, \\
Transition speed, Phase diagram}
\label{s_results}

\subsection{Alignment during the symmetric planar collision of two identical particles}
\label{subs_pair}

\begin{figure}
\begin{center}
\includegraphics[scale=0.333,angle=270]{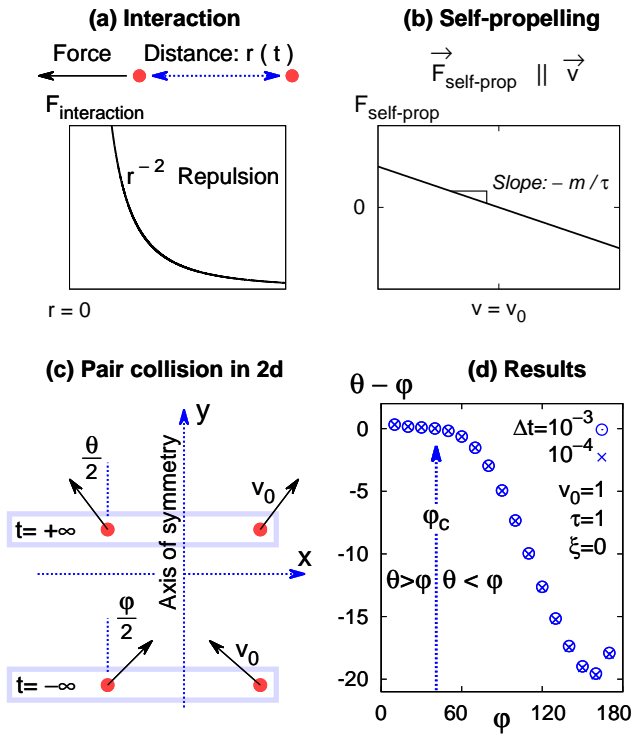}
\end{center}
\caption[]{
(Color online) Symmetric planar collision of 2 identical point-like particles according to Eq.\,(\ref{eq2}).
{\bf (a,b)} Illustration of the model. {\bf (c,d)} Geometry of the collision and results.
We integrate the equations of motion with the forward Euler method. 
Initial and final distances were $r=1,000$.
If the integration time step is reduced from $\Delta t=10^{-3}$ to $10^{-4}$,
the results change very little (see circles and crosses in panel d).
\label{figPair}
}
\end{figure}

Our goal is to see if Eq.\,(\ref{eq1}) leads to stable directional ordering in a system of many (interacting) particles.
A strong microscopic sign for macroscopic ordering can be 
the growth of the total momentum in a single collision of two particles.
%
%
To search for this microscopic behavior, we consider the simple scenario when two identical particles collide symmetrically in 2 dimensions without noise,
and they have the same speed, $v_0$, both long before and long after the collision.
Taking advantage of the collision's symmetry we follow only one of the two particles,
denote its path by ($x(t),y(t)$) and its velocity by ($v_x(t),v_y(t)$).
%
%
The collision's geometry -- together with the initial angle ($\varphi$) and the final angle ($\theta$) -- 
are defined in Fig.\,\ref{figPair}.
We set the $y$ axis to the symmetry axis of the collision and shift time ($t$) such that
at $t=0$ the two particles are closest to each other.
This implies also $v_x(0)=0$ and $y(0)=0$.
Note that during the collision $v_x(t)$ grows monotonously from $v_x(t=-\infty)<0$ to $v_x(t=+\infty)>0$.
%
%
For simplicity, we set now noise amplitudes to zero ($\xi=0$) \cite{noise}
and will return to nonzero noise ($\xi\ge 0$) only at Fig.\,\ref{figPow}.
In summary, the equations of motion of the analyzed particle ($x>0$) are

\bea
m\f{d v_x}{dt} & = & m\f{v_x}{v}\,\f{v_0-v}{\tau} + F[2x(t)] \, , \nonumber \\
m\f{d v_y}{dt} & = & m\f{v_y}{v}\,\f{v_0-v}{\tau}            \, . \label{eq2}
\eea

We set $v_0=1$ and $\tau=1$, and apply the repulsion term $F(r)=c\,r^{-2}$ with $c=1\,kg\,m^{3}\,s^{-2}$.
%
%
With these settings we find that two particles arriving at an angle of $\varphi<\varphi_c\approx 41$ degrees
depart at an angle ($\theta$) that is slightly above $\varphi$,
thus, the paths of the two particles become slightly less parallel during the collision.
However, for $\varphi>\varphi_c$ the two particles' paths become significantly {\it more parallel} during the collision ($\theta<\varphi$),
thus, the total momentum of the system grows (Fig.\,\ref{figPair}d).
Note that the repulsion term is conservative (i.e., derived from a potential),
thus, -- in contrast to the model of Ref.\,\cite{grossman} --
inealisticity is not necessary for this local growth of the momentum.
According to further simulations not shown here,
this critical angle, $\varphi_c$, exists for other repulsive interaction types as well.
For $F(r)=5[1+\sgn(1-r)]$ and $F(r)=10\exp(-r)$ we find the critical angles
$\varphi_c\approx 44$ (degrees) and $\varphi_c<5$ (degrees), respectively.

In the classical limit ($\tau\to\infty$) time is reversible, thus, the two particles' paths are both symmetric to the $y=0$ axis,
which gives $\theta=\varphi$.
In the overdamped limit ($\tau\to 0$) speeds are constant ($v_0$) and
the two equations in (\ref{eq2}) lead to a single equation (for $x\not= 0$):

\bea
\ddot{x}(t)\,x^2(t) = \bigg( 1 - [{\dot{x}}(t)/v_0]^2 \bigg) \, \f{c}{4m} \, .
\label{eq3}
\eea

\noindent
This equation is also reversible in time, and thus, in the overdamped limit, too, we have $\theta=\varphi$.
For finite values of $\tau$ there is no time reversal symmetry and usually
$\theta\not=\varphi$.

\subsection{Stable aligned motion of\\
many interacting particles}
\label{subs_coll}

We proceed with finite $\tau$ and zero noise. 
To test if the observed convergent pair collisions
cause a stable ordering of many particles,
we integrate numerically the motion of $N$ particles
in a rectangle with size $L$ and periodic boundaries.
We apply the midpoint method with time step $\Delta t$,
and for numerical efficiency we
cut off interactions at a particle-particle distance of $R$.
With these settings
we search for global directional ordering through
high ($E\approx 1$) values of the order parameter

\bea
E(t) = \f{1}{ N v_0 } \bigg| \sum_{i=1}^{N} \vvec_i(t) \bigg| \, .
\label{eqec}
\eea

\unitlength10mm
\begin{figure}
\begin{center}
\includegraphics[scale=0.37,angle=270]{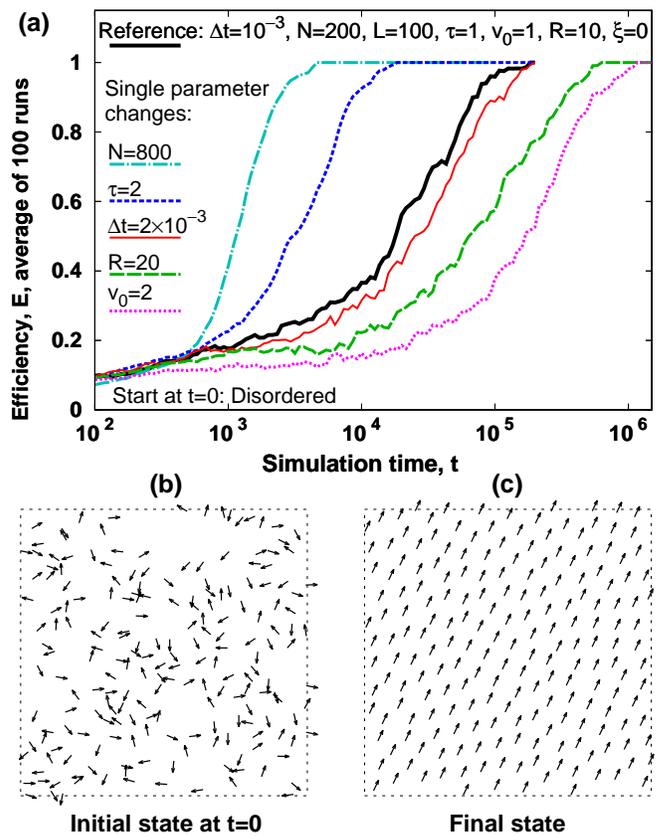}
\end{center}
\caption[]{
(Color online) Transition from disorder to order 
in a rectangle with periodic boundaries
and without noise ($\xi=0$).
At $t=0$ the system is started from the disordered state. 
During the simulations particles self-propel to maintain their speed
and they repel each other (see Eq.\,(\ref{eq1}) without noise).
{\bf (a)} Convergence to the $E\approx 1$ ordered state is
robust to changes of model parameters ($N$, $\tau$, $v_0$) 
and integration parameters ($R$, $\Delta t$).
{\bf (b)} Initial state: coordinates are random with distances above $L/(2\sqrt{N})$. 
Directions are random and speeds are $v_0$.
{\bf (c)} Final state: coordinates and directions are ordered.
\label{figTrans}
}
\end{figure}

With the parameter sets shown in Figure\,\ref{figTrans}a we find that particles started at {\it random initial} coordinates and
with random directions evolve to a stable state that has {\it globally ordered} velocities ($E\approx 1$) and coordinates.
Changes to the parameters of the model ($N$, $\tau$ or $v_0$)
or the integration ($R$ or $\Delta t$) do not affect the onset of this stable state,
only the time needed for the system to converge to it (Fig.\,\ref{figTrans}a).
In particular, for the analyzed finite values of $\tau$ convergence to the ordered state occurs in finite time.
As opposed to this finite time, there is no convergence to the ordered state in either the $\tau\to 0$ or the $\tau\to\infty$ limit
(recall that in both limiting cases time is reversible).
%
%
Thus, the time needed for ordering is minimal at a finite value of $\tau$.
%
Similarly, in Fig.\,\ref{figTrans}a we observe that the 
preferred speed of particles, $v_0$, also has a finite optimal value for fast ordering,
because (i) at $v_0=0$ there is no ordering and (ii) with growing $v_0$ convergence
to the ordered state becomes slower.
In contrast to this,
in time-discretized flocking models with explicit velocity alignment
the time needed for ordering is minimal not at a finite particle speed ($v_0$),
but in the $v_0\to\infty$ limit (with $v_0\Delta t/L=$const.),
because these models allow non-physically high accelerations when aligning velocities.
The current model avoids this artefact by using continuous time and not using explicit velocity alignment.

\subsection{Transition without noise:\\
Finite optimal density provides\\
shortest transition time}
\label{subs_dens}

\unitlength10mm
\begin{figure}
\begin{center}
\includegraphics[scale=0.37,angle=270]{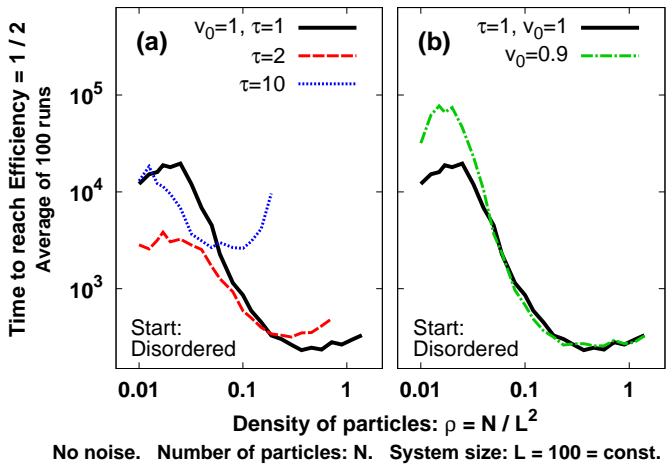}
\end{center}
\caption[]{
(Color online) The length of simulation time, $t$,
needed by the system to reach the average efficiency $\langle E\rangle = 1/2$
when running without noise ($\xi=0$) and started from the disordered state.
The results show averages over $100$ simulation runs.
The density of particles, $\rho=N/L^2$, is varied among data points.
In both panels
the black solid line was obtained with the reference parameter set shown in Fig.\,\ref{figTrans}a.
The two panels show results for various 
{\bf (a)} time constants of self-propelling, $\tau$, and
{\bf (b)} preferred velocities of motion, $v_0$.
We conclude that the most rapid convergence to the ordered state
occurs always at a finite density.
\label{fig_dens}
}
\end{figure}

In both natural systems and applications the speed
of convergence to an ordered state characterizes the stability of the ordered state.
This speed is often interpreted as the time necessary for making a collective decision \cite{conradt,ramseyer,neda,lee}.
Here we measure with several parameter sets of the model
the time necessary for the convergence of the self-driven particles
to the $E\approx 1$ ordered state.


In two-dimensional close packing
-- a hexagonal close packing of disks (filled circles) --
the portion of the total area covered is
$\pi/(2\sqrt{3})\approx 0.907$. 
In our case the $L\times L$ simulated area has side length $L=100$
and particle-particle interactions are cut off at a distance of $R=10$ of two particles.
Thus, approximately above the particle density $\rho_c=(R/L)^2\pi/(2\sqrt{3})\approx 0.009$
a typical particle on a typical simulation snapshot interacts with at least one other particle.
Consequently, in Fig.\,\ref{fig_dens} we analyze convergence to the ordered state at $\rho=0.01$ and above.
This figure shows the length of simulation time that the system needs to reach $E=1/2$ when started
from the disordered state without noise.
According to Fig.\,\ref{fig_dens}a (in agreement with Fig.\,\ref{figTrans}a)
for fast convergence there is an optimal density at each investigated parameter set.
This is in good agreement with the fundamental diagrams of 
unidirectional pedestrian motion \cite{chattaraj} and highway traffic \cite{chowdhury,helbing2}
showing an optimal density of the moving agents for maximal flow.
For a qualitative explanation consider the following.
At low density, biological or other agents (self-propelled particles)
may move in many different directions and
interact less frequently, thus they reach the ordered state in a longer time.
At medium density they interact more frequently
and confine each others' motion (due to their soft cores),
thus, they reach ordered motion more quickly.
Above medium density adding more particles
will not make the convergence to the ordered state faster.

Again, the decreasing speed of convergence
(to the directionally and spatially ordered state)
at high densities makes the current model more realistic
than time-discretized flocking models applying discontinuous 
(instantaneous) explicit velocity changes \cite{vicsek95,gregoire}.



\subsection{Transition with noise:\\
Critical slowing down and hysteresis}
\label{subs_noise}

\unitlength10mm
\begin{figure}
\begin{center}
\includegraphics[scale=0.365,angle=270]{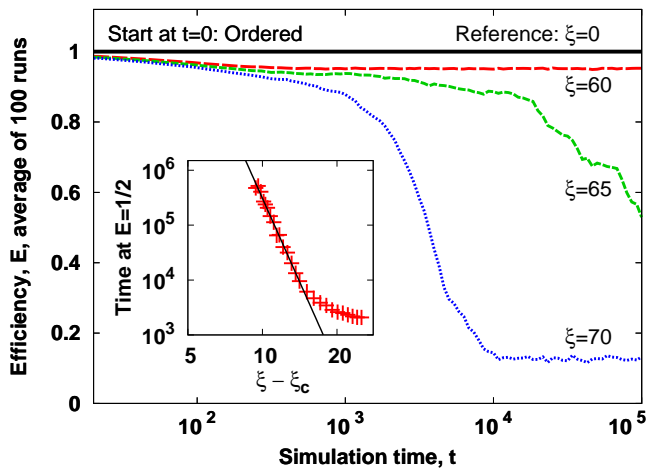}
\end{center}
\caption[]{
(Color online) Transition from order (Fig.\,\ref{figTrans}c) to disorder (Fig.\,\ref{figTrans}b).
At $t=0$ the system is in the ordered state.
The reference parameter set of Fig.\,\ref{figTrans}a is modified 
only by varying the noise amplitude, $\xi$.
The midpoint integration method is applied
by setting the length of the noise vector to 
$\xi\sqrt{\Delta t}$ and $\xi\sqrt{\Delta t/2}$ 
for the first and second step of an update, respectively.
{\bf Main panel.}
At low noise amplitudes the system remains in the ordered state,
and at high noise amplitudes it leaves the ordered state.
{\bf Inset.} The average efficiency reaches $1/2$ (from above) at a simulation time that
diverges -- as a function of the noise amplitude -- at a finite $\xi=\xi_c$ value similarly to a power-law.
The solid line is a fit to the $\xi=63\dots 68$ data points with the
parameters $\xi_c=53.9\pm 4.6$ (critical point) and $\gamma=-10.1\pm 4.1$ (exponent).
\label{figPow}
}
\end{figure}

To test the {\it stability of the ordered state},
we modify the reference parameter set (Fig.\,\ref{figTrans}a)
by applying nonzero noise amplitudes ($\xi>0$).
We find that the stability of the ordered steady state is lost at a finite (nonzero) noise amplitude (see Fig.\,\ref{figPow}).
In Figure\,\ref{figPow} note also
that the simulation time needed for the loss of stability diverges similarly to a power law as we approach this finite 
$\xi_c$ noise amplitude from above.
In other words, in the parameter range between the system's ordered state and disordered state we find a critical slowing down
indicating a dynamical phase transition \cite{hohenberg,start}.

\unitlength10mm
\begin{figure}
\begin{center}
\includegraphics[scale=0.365,angle=270]{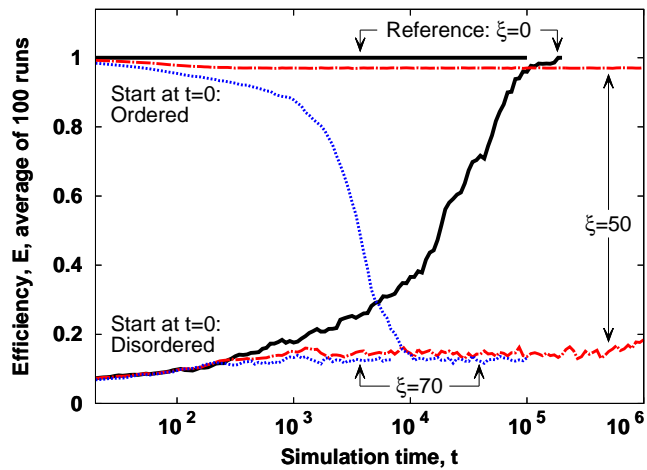}
\end{center}
\caption[]{
(Color online)
Searching for hysteresis.
We use the reference parameter values (Fig.\,\ref{figTrans}a)
or modify it by setting the noise amplitude. 
We find that at simulation time $t=10^6$
systems started from the ordered state ($E\approx 1$)
have a higher average efficiency 
than those started from the disordered state.
Thus, the system shows hysteresis with this parameter set and at this simulation time.
However, after longer simulations or with more particles
the two $\xi=50$ curves (marked) may converge to the same final average $E$ value.
\label{figX}
}
\end{figure}

In systems of self-propelled particles 
there has been a significant interest 
not only in the speed of the order-disorder transition, 
but also in the type of the transition \cite{vicsek95,gregoire,nagy07,aldana}.
To test the {\it if the transition observed here is discontinuous}, we search for hysteresis.
We start simulations from both states (ordered and disordered)
with the goal to allow the system to evolve to its final order parameter ($E$) value.
We find that with a noise amplitude of $\xi=50$ at simulation time $t=10^6$
the system's efficiency still strongly depends on its initial state (Fig.\,\ref{figX}).
This indicates hysteresis, i.e., a discontinuous transition, with the given parameter set and simulation time.
However, the system's slowing down shown in Fig.\,\ref{figPow} 
suggests that this numerically observed hysteresis may to completely disappear
both at longer simulation times and in the $N\to\infty$ limit.

\begin{figure}
\begin{center}
\includegraphics[scale=0.365,angle=270]{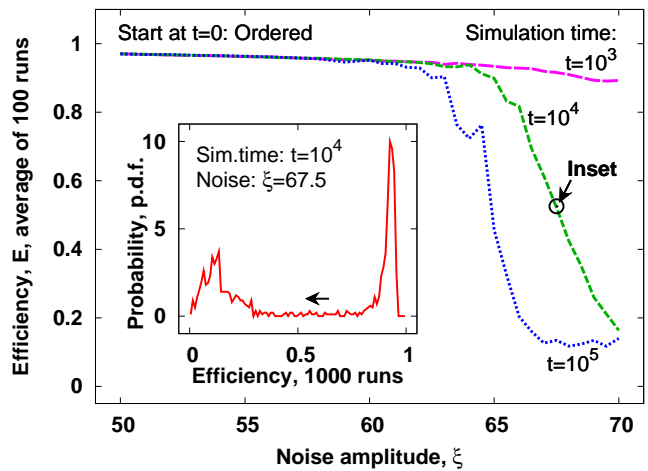}
\end{center}
\caption[]{
(Color online) Efficiency ($E$, ordering) as a function of noise amplitudes ($\xi$).
The reference parameter set of Fig.\,\ref{figTrans}a is changed by varying $\xi$.
{\bf Main panel.} 
With longer simulation times lower noise amplitudes are sufficient for 
replacing the ordered state with the disordered state.
A circle marks the parameter values used in the inset.
{\bf Inset.} 
At a finite simulation time and finite system size the bimodal distribution of the efficiencies of
$1\,000$ simulation runs shows the presence of hysteresis.
The peak of the high efficiency (ordered) state and the peak of the low efficiency (disordered) state do not move.
Only the few simulation runs (instances of the system) marked between the two peaks by a right-to-left arrow
are transitioning from the peak of the ordered state to the peak of the disordered state.
\label{figBimod}
}
\end{figure}

As an alternative to starting the system from both the ordered and the disordered state,
we search for {\it hysteresis} also by investigating the {\it distribution of order parameter} (efficiency) values.
We find that the critical noise amplitude at which the 
ordered state (Fig.\,\ref{figTrans}c)
is replaced by the disordered state (Fig.\,\ref{figTrans}b)
drops with increasing simulation time.
This is shown in the main panel of Fig.\,\ref{figBimod}.
We define the time of the transition as the time
when the system reaches half of the maximal efficiency:
$t(\langle E\rangle\approx 1/2)$.
With this definition
we observe (at the given finite time point) that at the time of the transition
most simulation instances are either ordered ($E\approx 1$)
or disordered, and only a few have intermediate efficiencies (Fig.\,\ref{figBimod}, inset).
This bimodal distribution is in good agreement with the results of Fig.\,\ref{figX}, 
and shows the presence of hysteresis.

\unitlength10mm

\subsection{Stability of phases and Phase diagram}
\label{subs_stabil}

\begin{figure}
\begin{center}
\includegraphics[scale=0.365,angle=270]{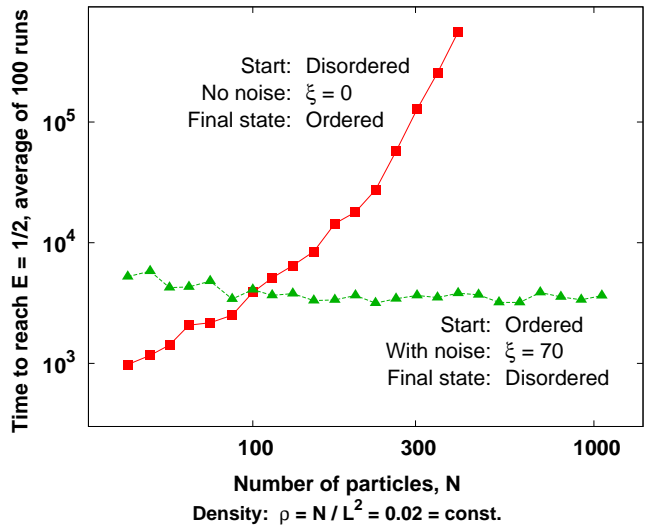}
\end{center}
\caption[]{
(Color online) Simulation time necessary for the disorder$\rightarrow$order 
transition without noise (red squares, solid line)
and the order$\rightarrow$disorder transition with noise (green triangles, dashed line).
The density of particles is constant.
With growing system size noise can destroy the ordered state in constant time (triangles).
On the other hand, in the absence of explicit noise ($\xi=0$)
the time needed for order to emerge from disorder 
grows at least as fast as a power law (squares).
We have tested that in the displayed system size range this 
speed of growth is below exponential.
\label{figStabi}
}
\end{figure}

Figures\,\ref{figX} and \ref{figBimod} show evidence for hysteresis in a finite system.
With growing system size and constant particle density we find that the hysteretic behavior becomes weaker.
While the ordered phase ``falls apart'' in constant time (at $\xi=70$) almost independently of the system's size,
the time needed for its re-emergence at $\xi=0$ grows rapidly with the number of particles (see Fig.\,\ref{figStabi}).
This behavior is a consequence of the local nature of the particles' interactions.
Once order is lost in a small region due to fluctuations,
the stabilizing effect of this region on neighbors is lost as well,
and return to the globally ordered state will be unlikely.
On the other side of the hysteresis loop, 
i.e., when the system is started from the disordered state with low noise,
the time that the locally ordered particle clusters will need
to merge to a single globally ordered cluster will grow rapidly
with the total number of particles.

Based on Fig.\,\ref{figStabi}
we see no evidence for the disappearance of the hysteresis loop at finite system 
sizes (only its weakening),
however, it seems to disappear entirely in the infinite system size limit.
Thus, we suggest that in the $N\to\infty$ limit and with constant density 
the critical noise amplitude, $\xi_c$ (see Fig.\,\ref{figPow}),
and the width of the hysteresis loop both converge to zero.
Moreover, based on Fig.\,\ref{fig_dens}, we suggest
that at a fixed system size ($L$) hysteresis is strongest at a finite density,
above which the width of the hysteresis loop and the critical noise amplitude both decrease.
In summary, we predict that at finite parameter values and also in 
the infinite system size (or infinite density) limit 
(i) both phases (ordered and disordered) are possible,
(ii) the ordered state is stable at small noise and the disordered state is stable at large noise.
We find a (iii) discontinuous (first order) transition with hysteresis at finite parameter values,
however, we predict that (iv) in the infinite limits there is no hysteresis,
and thus, the transition becomes continuous.

\unitlength10mm

\section{Discussion}

The current model can be derived as a limiting case of a
significantly simplified version of previous work 
by d'Orsogna et.al. on self-propelled particles with soft-core interactions \cite{dorsogna}.
To connect the two studies, first note that the current model applies a {\it minimal assumption} about controlling speed:
it does contain a first-order term, but omits the higher-order ($|v|^2\vec{v}$) viscous force applied by d'Orsogna et.al.
Second, Eq.\,\ref{eq2} contains no explicit attraction among the particles. 
This can be interpreted as a limiting case of the generalized Morse potential
of d'Orsogna et.al. reached by approaching zero strength or zero decay length of the attraction.
Recall also that the spatial confinement
necessary for the macroscopic ordering observed by us
is an indirect form of attraction acting mainly at the perimeters of the flock \cite{hamilton}.
To proceed with the comparison, note that in Figure 1 of Ref.\,\cite{dorsogna} the point of parameter space corresponding to the
attraction-free limiting case is on the borderline of regions with catastrophic (collapsing) behavior, however, 
when simulating Eq.\,(\ref{eq1}) we observe no collapse.

In addition to Ref.\,\cite{dorsogna} and the various forms of explicit velocity alignment,
further particle-based model components contributing to flocking include
attraction and adhesion\,\cite{couzin,dorsogna,johnston,strombom,szabo,mones,pimentel},
anisotropy\,\cite{ginelli,deseigne,peruani}
inelastic collisions\,\cite{grossman}
and coupled accelerations\,\cite{p.szabo}.
In particular, in models with explicit velocity alignment it was found
-- by a mapping to the majority voter model -- 
that extrinsic noise (also called: vectorial noise)
leads to a discontinuous transition \cite{pimentel}.
It may appear that the derivation in 
Ref.\,\cite{pimentel} explains the discontinuous transition observed in Fig.\,\ref{figX}.
However, the current model contains no explicit alignment (neighbor following), thus,
the mapping of Ref.\,\cite{pimentel} cannot be performed here,
which makes its subsequent proof about a discontinuous transition inapplicable for the current model.

We note also that our results are in agreement
with the recent finding in bacterial systems that the self-propulsion, 
rod shape, soft core repulsion and spatial confinement (which is an indirect form of a weak long-range attraction)
of the cells are sufficient for ordering \cite{peruani2}.
Interestingly, we find here that even less is sufficient,
because the rod shape -- i.e., anisotropy -- 
of the self-driven particles is not necessary for macroscopic ordering.

In summary, we find that in a fully (spatially and temporally) continuous model of collective motion
the following are {\it sufficient for stable global ordering}:
(a) self-propelling parallel to the current velocity, (b) radial repulsion and (c) spatial confinement.
Spatial confinement represents real obstacles (trees, walls, etc.) and flock perimeter avoidance by the participants \cite{hamilton}.
The absence of (fixed-time) explicit velocity alignment from the model avoids non-physically large accelerations.
We conclude that Eq.\,(\ref{eq1}) provides a {\it minimal and realistic description} of many flocking phenomena
involving insects, fish, birds, quadrupeds, humans or robots.

%
%
%


\begin{acknowledgements}
We thank T. Vicsek and P. Pollner
for discussions and computing facilities,
M. Nagy and A. Hein for comments and suggestions.
This project was supported by the 
EU ESF FuturICT.hu (T\'AMOP-4.2.2.C-11/1/KONV-2012-0013),
the Hungarian National Scientific Research Fund (OTKA NN 103114),
the National Natural Science Foundation of China (61202207 and 61472370) and the
China Postdoctoral Science Foundation (2012M520067).
\end{acknowledgements}


\begin{thebibliography}{9}
\bibitem{noise}
Even if the noise term is removed (i.e., $\xi=0$), rounding errors and the numerical integration technique add some noise to the results.
\bibitem{start}
To start a simulation at $t=0s$ from the spatially and directionally ordered state shown in Fig.\,\ref{figTrans}c,
we initialize particles at $t=-1,000s$ with fully aligned velocities and random coordinates (distances have a lower bound),
and then run the simulation until $t=0s$ to reach spatial ordering as well.
\end{thebibliography}
\end{document}